\documentclass[onecolumn,preprintnumbers,pra,hyperref]{revtex4}
\usepackage{amssymb}
\usepackage{amsfonts}
\usepackage{amsmath}
\usepackage{graphicx}

\begin{document}

\title{A new quantum mechanical photon counting distribution formula\thanks{Project
supported by the National Natural Science Foundation of China (Grant
Nos.10775097 and 10874174).}}
\author{Hong-yi Fan$^{1}$, Hong-chun Yuan$^{1}$\thanks{Corresponding author:
yuanhch@sjtu.edu.cn or yuanhch@126.com}, and Li-yun Hu$^{2}$\\$^{1}${\small Department of Physics, Shanghai Jiao Tong University, Shanghai
200240,} {\small \ China}\\$^{{\small 2}}${\small College of Physics and Communication Electronics,
Jiangxi Normal University, Nanchang 330022, China}}

\begin{abstract}
By virtue of density operator's P-representation in the coherent state
representation, we derive a new quantum mechanical photon counting
distribution formula. As its application, we find the photon counting
distribution for the pure squeezed state relates to the Legendre function,
which seems a new result.



\end{abstract}

\maketitle

\section{Introduction}

Photon counting by photoelectron devices is a very important topic in quantum
optics. Since it can judge the nonclassical features of light fields, and most
measurements of the electromagnetic field are based on the absorption of
photons via the photoelectric effect \cite{r0,r00}. So the problem of
photoelectric detection attracts the attention of many physicists and
scientists. Expression of quantum mechanical photon counting distribution was
first derived by Kelley and Kleiner \cite{r1}. For the case of single
radiation mode the probability distribution $\mathfrak{p}\left(  m,T\right)  $
of registering $m$ photoelectrons in the time interval $T$ is given
by\cite{r1,r2,r3}%

\begin{equation}
\mathfrak{p}\left(  m,T\right)  =Tr\left[  \rho \mathbf{\colon}\frac{\left(
\xi a^{\dagger}a\right)  ^{m}}{m!}e^{-\xi a^{\dagger}a}\colon \right]
,\label{1}%
\end{equation}
where $\rho$ is a single-mode density operator of the light field concerned,
$\mathbf{\colon \colon}$ stands for normal ordering, and $\xi$ is called the
quantum efficiency (a measure) of the detector. Mogilevtsev's group has as
well discussed no counts registered on a photonic detector \cite{r4}, i.e.,
for the case of $m=0$ in Eq.(\ref{1}). Recently, Fan and Hu \cite{1} derived
two formulas for photocounting; one involves $\rho$'s coherent state mean
value and the other is related to $\rho$'s Wigner function.

As an example, when $\rho \mathbf{=}\left \vert \alpha \right \rangle \left \langle
\alpha \right \vert $ is a pure coherent state, $\left \vert \alpha \right \rangle
=\exp \left(  \alpha a^{\dagger}-\alpha^{\ast}a\right)  \left \vert
0\right \rangle \equiv D\left(  \alpha \right)  \left \vert 0\right \rangle ,$ and
$a\left \vert \alpha \right \rangle =\alpha \left \vert \alpha \right \rangle $, then
the probability distribution is
\begin{equation}
\left \langle \alpha \right \vert \mathbf{:}\frac{\left(  \xi a^{\dagger
}a\right)  ^{m}}{m!}e^{-\xi a^{\dagger}a}\colon \left \vert \alpha \right \rangle
=\frac{\left(  \xi|\alpha|^{2}\right)  ^{m}e^{-\xi|\alpha|^{2}}}{m!}.\label{2}%
\end{equation}
which is the Poisson distribution as expected. Due to the operator identity
$e^{\lambda a^{\dagger}a}=\colon \exp[\left(  e^{\lambda}-1\right)  a^{\dagger
}a]\colon$\cite{2} and
\begin{equation}
e^{a^{\dagger}a\ln \left(  1-\xi \right)  }a^{m}e^{-a^{\dagger}a\ln \left(
1-\xi \right)  }=a^{m}\left(  1-\xi \right)  ^{-m},\label{3}%
\end{equation}
as well as%
\begin{equation}
a^{\dagger m}a^{m}=N\left(  N-1\right)  \cdots \left(  N-m+1\right)  ,\text{
}N=a^{\dagger}a.\label{3-1}%
\end{equation}
we see%
\begin{align}
\frac{\xi^{m}}{m!}\colon \left(  a^{\dagger}a\right)  ^{m}e^{-\xi a^{\dagger}%
a}\colon &  =\frac{\xi^{m}}{m!}a^{\dagger m}a^{m}\left(  1-\xi \right)
^{-m}e^{a^{\dagger}a\ln \left(  1-\xi \right)  }\nonumber \\
&  =\frac{\xi^{m}}{m!}\sum \limits_{n=0}^{\infty}\left \vert n\right \rangle
\left \langle n\right \vert N\left(  N-1\right)  \cdots \left(  N-m+1\right)
\left(  1-\xi \right)  ^{-m}e^{a^{\dagger}a\ln \left(  1-\xi \right)
}\nonumber \\
&  =\frac{\xi^{m}}{m!}\sum \limits_{n=m}^{\infty}\left \vert n\right \rangle
\left \langle n\right \vert n\left(  n-1\right)  \cdots \left(  n-m+1\right)
\left(  1-\xi \right)  ^{n-m}\nonumber \\
&  =\sum \limits_{n=m}^{\infty}\binom{n}{m}\xi^{m}\left(  1-\xi \right)
^{n-m}\left \vert n\right \rangle \left \langle n\right \vert ,\label{4}%
\end{align}
where $\left \vert n\right \rangle =\frac{a^{\dagger n}}{\sqrt{n!}}\left \vert
0\right \rangle $ is the number state. Substituting Eq.(\ref{4}) into
Eq.(\ref{1}) yields%
\begin{equation}
\mathfrak{p}\left(  m,T\right)  =\sum_{n=m}^{\infty}\mathcal{P}_{n}\binom
{n}{m}\xi^{m}\left(  1-\xi \right)  ^{n-m},\label{5}%
\end{equation}
where $\mathcal{P}_{n}=\left \langle n\right \vert \rho \left \vert n\right \rangle
.$ Eq.(\ref{5}) is named as Bernoulli distribution in particle-number
representation \cite{r2,r3}. However, for many density operators other than
the pure coherent state, directly using Eq.(\ref{1}) is difficult and is
inconvenient, so in this work we shall reform it.

\section{New photon counting distribution formula derived by density
operator's P-representation}

Now we instead use density operator's P-representation\cite{1a}
\begin{equation}
\rho \mathbf{=}\int d^{2}\alpha P\left(  \alpha \right)  \left \vert
\alpha \right \rangle \left \langle \alpha \right \vert \label{6}%
\end{equation}
to analyse Eq. (\ref{1}), $\left \vert \alpha \right \rangle $ is the coherent
state. Substituting Eq.(\ref{6}) into Eq.(\ref{1}) yields
\begin{align}
\mathfrak{p}\left(  m,T\right)   &  =\int d^{2}\alpha P\left(  \alpha \right)
\left \langle \alpha \right \vert \mathbf{:}\frac{\left(  \xi a^{\dagger
}a\right)  ^{m}}{m!}e^{-\xi a^{\dagger}a}\colon \left \vert \alpha \right \rangle
\nonumber \\
&  =\int d^{2}\alpha P\left(  \alpha \right)  \frac{\left(  \xi|\alpha
|^{2}\right)  ^{m}}{m!}e^{-\xi|\alpha|^{2}}. \label{7}%
\end{align}
Using $P\left(  \alpha \right)  ^{\prime}$s expression\cite{1b}%
\begin{equation}
P\left(  \alpha \right)  =e^{|\alpha|^{2}}\int \frac{d^{2}\beta}{\pi
}\left \langle -\beta \right \vert \rho \left \vert \beta \right \rangle
e^{|\beta|^{2}+\alpha \beta^{\ast}-\beta \alpha^{\ast}}, \label{8}%
\end{equation}
where $\left \vert \beta \right \rangle $ is also a coherent state, Eq. (\ref{7})
becomes%
\begin{align}
\mathfrak{p}\left(  m,T\right)   &  =\frac{\xi^{m}}{m!}\int \frac{d^{2}\beta
}{\pi}e^{|\beta|^{2}}\left \langle -\beta \right \vert \rho \left \vert
\beta \right \rangle \int d^{2}\alpha|\alpha|^{2m}e^{\left(  1-\xi \right)
|\alpha|^{2}+\alpha \beta^{\ast}-\beta \alpha^{\ast}}\nonumber \\
&  =\frac{\xi^{m}}{\left(  \xi-1\right)  ^{m+1}}\int \frac{d^{2}\beta}{\pi
}\left \langle -\beta \right \vert \rho \left \vert \beta \right \rangle e^{\frac
{\xi-2}{\xi-1}|\beta|^{2}}L_{m}\left(  \frac{|\beta|^{2}}{\xi-1}\right)  ,
\label{9}%
\end{align}
where $L_{m}(x)$ is the Laguerre polynomial,%
\begin{equation}
L_{m}\left(  x\right)  =\sum_{l=0}^{m}(-1)^{l}\dbinom{m}{l}\frac{x^{l}}%
{l!}=\sum_{l^{\prime}=0}^{m}(-1)^{m-l^{\prime}}\dbinom{m}{m-l^{\prime}}%
\frac{x^{m-l^{\prime}}}{\left(  m-l\right)  !} \label{11}%
\end{equation}
and we have used
\begin{equation}
\int \frac{d^{2}z}{\pi}z^{n}z^{\ast m}e^{\epsilon|z|^{2}+Bz+Cz^{\ast}%
}=e^{-BC/A}\sum_{l=0}^{\min[m,n]}\frac{n!m!B^{m-l}C^{n-l}}%
{l!(n-l)!(m-l)!\left(  -\epsilon \right)  ^{n+m-l+1}},\operatorname{Re}%
\epsilon<0. \label{10}%
\end{equation}
Eq.(\ref{9}) is a new formula, different from that in Ref.\cite{1}.

\section{Applications}

As some applications of Eq.(\ref{9}), substituting the pure coherent state
$\rho \mathbf{=}\left \vert \alpha \right \rangle \left \langle \alpha \right \vert $
into Eq.(\ref{9}), we have%
\begin{align}
\mathfrak{p}_{\left \vert \alpha \right \rangle \left \langle \alpha \right \vert
}\left(  m,T\right)   &  =\frac{\xi^{m}}{\left(  \xi-1\right)  ^{m+1}}%
\int \frac{d^{2}\beta}{\pi}\left \langle -\beta \right \vert \left.
\alpha \right \rangle \left \langle \alpha \right.  \left \vert \beta \right \rangle
e^{\frac{\xi-2}{\xi-1}|\beta|^{2}}L_{m}\left(  \frac{|\beta|^{2}}{\xi
-1}\right)  \nonumber \\
&  =\frac{\xi^{m}e^{-|\alpha|^{2}}}{\left(  \xi-1\right)  ^{m+1}}\int
\frac{d^{2}\beta}{\pi}e^{\frac{-|\beta|^{2}}{\xi-1}+\beta \alpha^{\ast}%
-\alpha \beta^{\ast}}L_{m}\left(  \frac{|\beta|^{2}}{\xi-1}\right)  ,\label{c1}%
\end{align}
where $\left \langle \alpha \right.  \left \vert \beta \right \rangle =\exp \left[
-\frac{1}{2}\left(  |\alpha|^{2}+|\beta|^{2}\right)  +\alpha^{\ast}%
\beta \right]  .$ Enlightened by the generating function of the Laguerre
polynomial\cite{b1}
\begin{equation}
\sum_{n=0}^{\infty}L_{n}\left(  x\right)  t^{n}=\left(  1-t\right)  ^{-1}%
\exp \left(  \frac{-xt}{1-t}\right)  ,\label{c2}%
\end{equation}
so%
\begin{equation}
L_{n}\left(  x\right)  =\frac{1}{n!}\frac{\partial^{n}}{\partial t^{n}}\left.
\left[  \left(  1-t\right)  ^{-1}\exp \left(  \frac{-xt}{1-t}\right)  \right]
\right \vert _{t=0},\label{c3}%
\end{equation}
we rewrite Eq.(\ref{c1}) as%
\begin{align}
\mathfrak{p}_{\left \vert \alpha \right \rangle \left \langle \alpha \right \vert
}\left(  m,T\right)   &  =\frac{\xi^{m}e^{-|\alpha|^{2}}}{m!\left(
\xi-1\right)  ^{m+1}}\frac{\partial^{m}}{\partial t^{m}}\left.  \int
\frac{d^{2}\beta}{\left(  1-t\right)  \pi}e^{\frac{1}{\left(  \xi-1\right)
\left(  t-1\right)  }|\beta|^{2}+\alpha^{\ast}\beta-\alpha \beta^{\ast}%
}\right \vert _{t=0}\nonumber \\
&  =\frac{\xi^{m}e^{-\xi \left \vert \alpha \right \vert ^{2}}}{m!\left(
\xi-1\right)  ^{m}}\frac{\partial^{m}}{\partial t^{m}}\left.  e^{\allowbreak
t\left(  \xi-1\right)  \left \vert \alpha \right \vert ^{2}}\right \vert
_{t=0}\nonumber \\
&  =\frac{\xi^{m}\left \vert \alpha \right \vert ^{2m}e^{-\xi \left \vert
\alpha \right \vert ^{2}}}{m!},\label{c4}%
\end{align}
where we have used the formula%

\begin{equation}
\int \frac{d^{2}z}{\pi}e^{\epsilon|z|^{2}+Bz+Cz^{\ast}}=-\frac{e^{-BC/\epsilon
}}{\epsilon},\operatorname{Re}\epsilon<0.\label{c5}%
\end{equation}
It is clear that Eq.(\ref{c4}) exactly agrees with Eq.(\ref{2}). Comparing
Eqs.(\ref{c1}) with Eq.(\ref{c4}) implies a new integration formula%
\begin{equation}
\int \frac{d^{2}\beta}{\pi}L_{m}\left(  \frac{|\beta|^{2}}{\xi-1}\right)
e^{\frac{-|\beta|^{2}}{\xi-1}+\beta \alpha^{\ast}-\alpha \beta^{\ast}}%
=\frac{\left(  \xi-1\right)  ^{m+1}\left \vert \alpha \right \vert ^{2m}}%
{m!}e^{\left(  1-\xi \right)  \left \vert \alpha \right \vert ^{2}}.\label{c6}%
\end{equation}
Especially, when $\alpha=0,$ in Eq.(\ref{c6}) only $m=0$ term survives, so we
obtain%
\begin{equation}
\int \frac{d^{2}\beta}{\pi}e^{-|\beta|^{2}}L_{m}\left(  |\beta|^{2}\right)
=1.\label{c7}%
\end{equation}

Now we calculate another density operator
\begin{equation}
\rho_{c}\mathbf{=}\left(  1-e^{-f}\right)  e^{-fa^{\dagger}a} \label{17}%
\end{equation}
with $f=\frac{\omega \hbar}{KT}$, which represents a filtered one-mode chaotic
light. Due to%
\begin{equation}
\left(  1-e^{-f}\right)  \left \langle -\beta \right \vert \colon e^{\left(
e^{-f}-1\right)  a^{\dagger}a}\colon \left \vert \beta \right \rangle =\left(
1-e^{-f}\right)  e^{-\left(  1+e^{-f}\right)  |\beta|^{2}}, \label{17-1}%
\end{equation}
substituting Eq.(\ref{17-1}) into Eq.(\ref{9}), we have
\begin{align}
\mathfrak{p}_{c}\left(  m,T\right)   &  =\frac{\xi^{m}\left(  1-e^{-f}\right)
}{\left(  \xi-1\right)  ^{m+1}}\int \frac{d^{2}\beta}{\pi}L_{m}\left(
\frac{|\beta|^{2}}{\xi-1}\right)  e^{-\left(  1+e^{-f}-\frac{\xi-2}{\xi
-1}\right)  |\beta|^{2}}\nonumber \\
&  =\left(  1-e^{-f}\right)  \left(  \frac{\xi}{\xi-1}\right)  ^{m}\int
_{0}^{\infty}drL_{m}\left(  r\right)  e^{-\left(  1+\xi e^{-f}-e^{-f}\right)
r}. \label{19}%
\end{align}
With the help of the formula\cite{b2}%
\begin{equation}
\int_{0}^{\infty}drL_{m}\left(  r\right)  e^{-gr}=\frac{\left(  g-1\right)
^{m}}{g^{m+1}}, \label{20}%
\end{equation}
Equation (\ref{19}) becomes%
\begin{equation}
\mathfrak{p}_{c}\left(  m,T\right)  =\frac{\left(  e^{f}-1\right)  \xi^{m}%
}{\left(  e^{f}+\xi-1\right)  ^{m+1}}. \label{21}%
\end{equation}
Noting $\left(  e^{f}-1\right)  ^{-1}=\left(  e^{\frac{\omega \hbar}{kT}%
}-1\right)  ^{-1}\equiv$ $\mathsf{\bar{n},}$ representing\ the Bose-Einstein
statistics, Eq.(\ref{21}) leads to the photon counting probability for the
chaotic light%
\begin{equation}
\mathfrak{p}_{c}\left(  m,T\right)  =\frac{\left(  \xi \mathsf{\bar{n}}\right)
^{m}}{\left(  1\mathsf{+\xi \bar{n}}\right)  ^{m+1}}, \label{22}%
\end{equation}
which agrees with the known result, so our approach's correctness is confirmed.

When $\rho_{s}$ is a pure squeezed state%
\begin{equation}
\rho_{s}=\sec \text{h}\lambda e^{\frac{\tanh \lambda}{2}a^{\dagger2}}\left \vert
0\right \rangle \left \langle 0\right \vert e^{\frac{\tanh \lambda}{2}a^{2}%
},\label{23}%
\end{equation}
where $\lambda$ is the squeezing parameter, after $\rho_{s}$ being substituted
into Eq.(9) we see%
\begin{equation}
\mathfrak{p}_{s}\left(  m,T\right)  =\frac{\xi^{m}}{\left(  \xi-1\right)
^{m+1}}\sec \text{h}\lambda \int \frac{d^{2}\beta}{\pi}e^{-\frac{|\beta|^{2}}%
{\xi-1}+\left(  \beta^{\ast2}+\beta^{2}\right)  \frac{\tanh \lambda}{2}}%
L_{m}\left(  \frac{|\beta|^{2}}{\xi-1}\right)  .\label{24}%
\end{equation}
Then using Eq.(\ref{c3}) and the integral formula\cite{b3}%
\begin{align}
&  \int \frac{d^{2}z}{\pi}\exp \left(  \zeta \left \vert z\right \vert ^{2}+\xi
z+\eta z^{\ast}+fz^{2}+gz^{\ast2}\right)  \nonumber \\
&  =\frac{1}{\sqrt{\zeta^{2}-4fg}}\exp \left[  \frac{-\zeta \xi \eta+\xi
^{2}g+\eta^{2}f}{\zeta^{2}-4fg}\right]  ,\label{25}%
\end{align}
with Re$\left(  \zeta+f+g\right)  <0,\ $Re$\left(  \frac{\zeta^{2}-4fg}%
{\zeta+f+g}\right)  <0$ or Re$\left(  \zeta-f-g\right)  <0,$ Re$\left(
\frac{\zeta^{2}-4fg}{\zeta-f-g}\right)  <0,$ we have\bigskip%
\begin{align}
\mathfrak{p}_{s}\left(  m,T\right)   &  =\frac{\xi^{m}\sec \text{h}\lambda
}{m!\left(  \xi-1\right)  ^{m+1}}\frac{\partial^{m}}{\partial t^{m}}\left.
\int \frac{d^{2}\beta}{\left(  1-t\right)  \pi}e^{\left(  \beta^{\ast2}%
+\beta^{2}\right)  \tanh \lambda/2-\frac{1}{\left(  \xi-1\right)  \left(
1-t\right)  }|\beta|^{2}}\right \vert _{t=0}\nonumber \\
&  =\frac{\xi^{m}\sec \text{h}\lambda}{m!\left(  \xi-1\right)  ^{m}}%
\frac{\partial^{m}}{\partial t^{m}}\left.  \frac{1}{\sqrt{1-\left(
1-t\right)  ^{2}G^{2}}}\right \vert _{t=0},\label{26}%
\end{align}
where%
\begin{equation}
G^{2}\equiv \left(  \xi-1\right)  ^{2}\tanh^{2}\lambda.\label{27}%
\end{equation}
Using the generating function of Legendre polynomial\cite{b1}%
\begin{equation}
\left(  1-2xt+t^{2}\right)  ^{-1/2}=\sum_{n=0}^{\infty}P_{n}\left(  x\right)
t^{n}\label{29}%
\end{equation}
or%
\begin{equation}
P_{n}\left(  x\right)  =\frac{1}{n!}\frac{\partial^{n}}{\partial t^{n}}\left.
\frac{1}{\sqrt{1-2xt+t^{2}}}\right \vert _{t=0},\label{30}%
\end{equation}
finally we have%

\begin{equation}
\mathfrak{p}_{s}\left(  m,T\right)  =\frac{\xi^{m}\sec \text{h}\lambda \tanh
^{m}\lambda}{\left(  1-G^{2}\right)  ^{1/2}\left(  G^{2}-1\right)  ^{m/2}%
}P_{m}\left(  \frac{G}{\sqrt{G^{2}-1}}\right)  \label{31}%
\end{equation}
which relates to the Legendre function, and this is a new result.

For the last example, when $\rho_{d}$ is the displaced chaotic field
\begin{equation}
\rho_{d}=\left(  1-e^{-f}\right)  D\left(  \alpha \right)  e^{-fa^{\dagger}%
a}D^{-1}\left(  \alpha \right)  \label{32}%
\end{equation}
with $D\left(  \alpha \right)  =\exp \left(  \alpha a^{\dag}-\alpha^{\ast
}a\right)  $, we convert $\rho_{d}$ into its normal ordering form\cite{b4}%
\begin{equation}
\rho_{d}=\frac{1}{\bar{n}+1}\colon \exp \left[  \frac{-\left(  \alpha-a\right)
\left(  \alpha^{\ast}-a^{\dag}\right)  }{\bar{n}+1}\right]  \colon \label{33}%
\end{equation}
and substitute it into Eq.(\ref{9}) yields%

\begin{align}
&  \mathfrak{p}_{d}\left(  m,T\right)  =\frac{\xi^{m}e^{-\frac{\left \vert
\alpha \right \vert ^{2}}{\bar{n}+1}}}{m!\left(  \bar{n}+1\right)  \left(
\xi-1\right)  ^{m+1}}\frac{\partial^{m}}{\partial t^{m}}\left.  \int
\frac{d^{2}\beta}{\left(  1-t\right)  \pi}e^{\frac{\bar{n}\xi+\bar{n}t-\bar
{n}t\xi+1}{\left(  \xi-1\right)  \left(  \bar{n}+1\right)  \left(  t-1\right)
}\allowbreak \left \vert \beta \right \vert ^{2}+\frac{\alpha^{\ast}}{\bar{n}%
+1}\beta-\frac{\alpha}{\bar{n}+1}\beta^{\ast}}\right \vert _{t=0}\nonumber \\
&  =\frac{\xi^{m}e^{-\frac{\xi \left \vert \alpha \right \vert ^{2}}{1+n\xi}}%
}{m!\left(  \xi-1\right)  ^{m}}\frac{\partial^{m}}{\partial t^{m}}\left.
\frac{\frac{1}{\bar{n}\xi+1}}{1-\frac{\bar{n}\left(  \xi-1\right)  }{\bar
{n}\xi+1}t}\exp \left[  \frac{\frac{\xi-1}{\left(  n\xi+1\right)  ^{2}%
}\left \vert \alpha \right \vert ^{2}t}{1-\frac{\bar{n}\left(  \xi-1\right)
}{\bar{n}\xi+1}t}\right]  \right \vert _{t=0}\nonumber \\
&  =\frac{\left(  \bar{n}\xi \right)  ^{m}e^{-\frac{\xi \left \vert
\alpha \right \vert ^{2}}{1+n\xi}}}{\left(  \bar{n}\xi+1\right)  ^{m+1}}%
L_{m}\left[  \frac{-\left \vert \alpha \right \vert ^{2}}{\bar{n}\left(
n\xi+1\right)  }\right]  , \label{34}%
\end{align}
where we have considered Eqs.(\ref{c3}) and (\ref{c5}). It is interesting to
see that when $\xi=1,\bar{n}=-1/2,$ $\rho_{d}$ in Eq.(\ref{33}) becomes
$\rho_{d}=2\colon \exp \left[  -2\left(  \alpha-a\right)  \left(  \alpha^{\ast
}-a^{\dag}\right)  \right]  \colon=2\pi \Delta(\alpha),$ where $\Delta(\alpha)$
is the Wigner operator. Then using $\mathbf{\colon}e^{-a^{\dagger}a}%
\colon=\left \vert 0\right \rangle \left \langle 0\right \vert ,$ $\left \vert
m\right \rangle =\frac{a^{\dagger m}}{\sqrt{m!}}\left \vert 0\right \rangle $, we have%

\[
\mathfrak{p}_{d}\left(  m,T\right)  =2\pi \left \langle m\right \vert
\Delta(\alpha)\left \vert m\right \rangle =2\left(  -1\right)  ^{m}%
e^{-2\left \vert \alpha \right \vert ^{2}}L_{m}\left[  4\left \vert \alpha
\right \vert ^{2}\right]  \equiv2\pi W\left(  \alpha \right)  ,
\]
where $W\left(  \alpha \right)  $ is just the Wigner function of the number
state $\left \vert m\right \rangle $\cite{b5}$.$ This is a\ good check for the
correctness of Eq.(\ref{34}).

In summary, by virtue of density operator's P-representation in the coherent
state representation, we derive a new quantum mechanical photon counting
distribution formula. As its application, we find that the photon counting
distribution for the pure squeezed state relates to the Legendre function.

\end{document}